\documentclass{appolb}
\usepackage{epsfig}
\newcommand{\lsim}{\raisebox{-0.13cm}{~\shortstack{$<$ \\[-0.07cm] $\sim$}}~}
\newcommand{\gsim}{\raisebox{-0.13cm}{~\shortstack{$>$ \\[-0.07cm] $\sim$}}~}
\def\s{{\cal S}}
\def\tr{{\rm Tr}}
\begin{document}
\pagestyle{plain}
\title{TEV STRINGS AND \\ ULTRAHIGH-ENERGY COSMIC RAYS%
\thanks{Work supported by MCYT, Junta de Andaluc{\'\i}a and the
European Union under contracts FPA2000-1558, FQM-101 and HPRN-CT-2000-00149,
respectively.}%
}
\author{Jos\'e I. Illana
\footnote{
	Presented at the XXV School of Theoretical Physics, 
        Ustro\'n, Poland, 9--16 Sept 2001.}
\address{
  Centro Andaluz de F{\'\i}sica de las Part{\'\i}culas Elementales (CAFPE)
  and \\ Departamento de F{\'\i}sica Te\'orica y del Cosmos,
	 Universidad de Granada, \\ 
	 E-18071 Granada, Spain}}
\maketitle
\vspace{-6.5cm}
\begin{flushright}
CAFPE-5/01\\
UG-FT-135/01
\end{flushright}
\vspace{4.5cm}

\begin{abstract}
The origin and nature of ultrahigh-energy cosmic ray events, 
above the Greisen-Zatsepin-Kuzmin (GZK) cutoff energy, 
constitute a long-standing, unsolved mistery. 
Neutrinos are proposed candidates but their standard interactions 
with matter are too weak.
In the context of a TeV-scale string theory, 
motivated by possible extra space dimensions,
the neutrino-nucleon scattering is examined.
Resonant string contributions increase substantially the standard
model neutrino-nucleon cross section. 
Although they seem insufficient to explain the trans-GZK cosmic 
ray events, their effects might be detected in next experiments. 
\end{abstract}

\section{The mistery of ultrahigh-energy cosmic rays}

Cosmic rays (CRs) were discovered in balloon experiments by 
V. Hess in 1911. They have always been a fruitful
particle physics laboratory: positrons, muons, pions and many other
particles were first observed there. CRs have a broad spectrum, 
ranging from 1 GeV to $10^{20}$ eV, including energies far beyond 
the ones available at man-made accelerators. The flux of CRs decreases
rapidly with the energy: \eg 1 particle/m$^2$/s at $10^{11}$ eV and only
1 particle/km$^2$/year at about $10^{18}$ eV.

At high energies, a CR may hit an atmospheric nucleon producing a 
cascade of secondary particles, extensive air showers (EAS),
that cover at ground an area with a diameter of tens of meters at
$10^{14}$ eV to several kilometers at $10^{20}$ eV. 
Incidentally, it is easier to observe the EAS at ground level
for energies where direct detection of CRs at satellite 
or balloon experiments becomes impractical.
The {\em primary} particle (mass and energy) can only be inferred 
from the parameters of the shower, such as number of electrons, muons 
or hadrons, shapes of lateral distributions, etc.

We will focus on the part of the spectrum above the Greisen-Zatsepin-Kuzmin
cutoff energy \cite{gzk} $E_{\rm GZK}\approx 5\times 10^{19}$ 
eV (see below). Unexpectedly, over seventy trans-GZK {\em ultrahigh-energy 
cosmic ray} events have been observed over the last forty years by five 
different experiments \cite{uhecr}. These events exhibit large-scale 
isotropy and small-scale anisotropy (a very unlikely pairing within 
the angular resolution occurs: three doublets and one triplet!).
They constitute a puzzle in modern astrophysics. The main open
questions are:

\begin{itemize}

\item
{\em What is the primary particle?} 

{\em Protons} with energies above the reaction threshold $E_{\rm GZK}$ 
lose part of their energy by scattering with cosmic microwave background
(CMB) photons:
$$
p+\gamma_{2.7{\rm K}}\to\Delta^+\to p+\pi^0\ (n+\pi^+).
$$
Above a distance of $D_{\rm GZK}\approx 50$ Mpc
the energy of the proton gets below $10^{20}$ eV, regardless of 
its initial value, after a series of such collisions.
This implies that the source cannot be too distant for trans-GZK events
if they are originated by protons.
Similar cutoffs exist at lower energies for other primaries:
{\em nuclei} undergo photodisintegration in CMB and IR radiations;
{\em electrons} lose energy very rapidly via synchrotron radiation; and
{\em photons} have a relatively short absorption length.

\item
{\em What is the acceleration mechanism?}

The conventional (Fermi) mechanism is the stochastic shock-wave acceleration 
in magnetized clouds. Powerful enough accelerators \cite{hillas}
are pulsars, AGN cores
and jets or radio-loud quasars. Gamma ray bursters may accelerate protons 
up to $10^{20}$ eV but they are too distant objects \cite{grb}. Alternatively, 
``top-down'' scenarios \cite{top-down} have been proposed, based on the 
decay or annihilation of super-heavy particles produced by topological 
defects \cite{topdef} or supermassive cosmological relics \cite{cosrel}.
Another possibility is tha assumption of exotic physical laws, such as 
breakdowns of Lorentz invariance \cite{lorinv} or general relativity 
\cite{genrel}.

\item
{\em Where are the sources?}

According to the conventional acceleration mechanism there are very 
few sources {\em at sight} within the GZK distance, insufficient
to explain the number of observed trans-GZK events.
Directional and temporal decorrelation with the source due to magnetic 
fields of the order of $\mu G$ has been proposed as a possible
explanation \cite{Farrar:2000bw}. Distant sources like BL Lacertae
blazers, outside the GZK volume, have received attention recently, since
they may emit very high energy photons cascading into secondary ones 
still above the GZK cutoff while approaching us \cite{Kalashev:2001qp}.

\end{itemize}

To evade most of the problems above, {\em neutrinos} are good candidates:
they are {\em neutral} (unbent by magnetic fields, allowing pairing),
{\em stable} and able to {\em propagate unimpeded to earth} from very 
distant sources: even if their cross section was of hadronic size, their
mean free path would be of the order of 100 Gpc among galaxies and 1
Mpc within a galaxy, so they can only be stopped by the atmosphere.
It is also suggestive that the flux of cosmogenic 
neutrinos matches well the flux of observed trans-GZK CRs, since the 
energy-degrading protons photoproduce pions decaying into neutrinos.
The main problem is that neutrino cross sections are smaller than those of 
hadronic or electromagnetic interactions by six orders of magnitude.
Two possible types of solutions have been proposed to the small cross section
problem of neutrinos:

-- ``$Z$-bursts'' \cite{z-burst}. 
   Cosmogenic neutrinos are not the primaries but resonantly create a 
   local flux of nucleons and photons (within the GZK distance) 
   with $E\gsim E_{\rm GZK}$ via scattering with the cosmic neutrino
   background. The resonance is a $Z$ boson. The necessary neutrino mass 
   range fits well with the one inferred from oscillation experiments,
   but large neutrino fluxes are required.
   Similarly, ``gravi-bursts'' \cite{gravi-burst} may proceed via
   resonant Kaluza-Klein excitations of
   the graviton in theories with large compact extra dimensions or in
   localized gravity models.

-- Enhance the neutrino cross section at high energies.
   Neutrinos are here the primary particles reaching the 
   atmosphere. This will be our approach, in the context of a
   weakly-coupled string theory assuming extra space dimensions
   at the TeV scale.

\section{The effects of extra dimensions}

Extensions of the Standard Model (SM) with extra dimensions 
offer new ways to accommodate the hierarchies observed 
in particle physics \cite{Antoniadis:1990ew}. A very attractive
possibility would be to bring the scale of unification 
with gravity from $M_{\rm Planck}\approx 10^{19}$ GeV down to the 
electroweak scale $M_{\rm EW}\approx 1$ TeV. This 
could result if gravity propagates along 
a (4+$n$)--dimensional flat space with $n$ compact 
submillimeter dimensions (ADD model \cite{Arkani-Hamed:1998rs}) 
or along a (4+1)--dimensional slice of anti-deSitter space with a warp
factor in the metric (RS model \cite{Randall:1999ee}).
These higher dimensional field theories, however, must be 
considered effective low-energy limits only valid below the
mass scale of a more fundamental theory. And nowadays,
only string theory \cite{pol98} provides a 
consistent framework for the unification of gravity with the 
standard model. 

In string theory the massless graviton comes as the zero mode of a
closed string, whereas the gauge bosons and matter fields are the 
lightest modes of open strings. The string vibration modes, with 
squared masses $M^2=n M^2_S$ (integer $n>1$), are the so called string
Regge (SR) excitations. The fundamental string scale is 
$M_S=\alpha'^{-1/2}$ where $\alpha'$ is the string tension.

At energies where the effects of a higher dimensional graviton are 
unsuppressed one expects the presence of its SR excitations giving 
an effect of the same size. This is necessary in order to avoid 
the pathologies of spin-2 field theories, such as the quantum field 
theory of gravity. 

\begin{figure}[htb]
\begin{center}
\begin{tabular}{lll}
(a) \epsfig{file=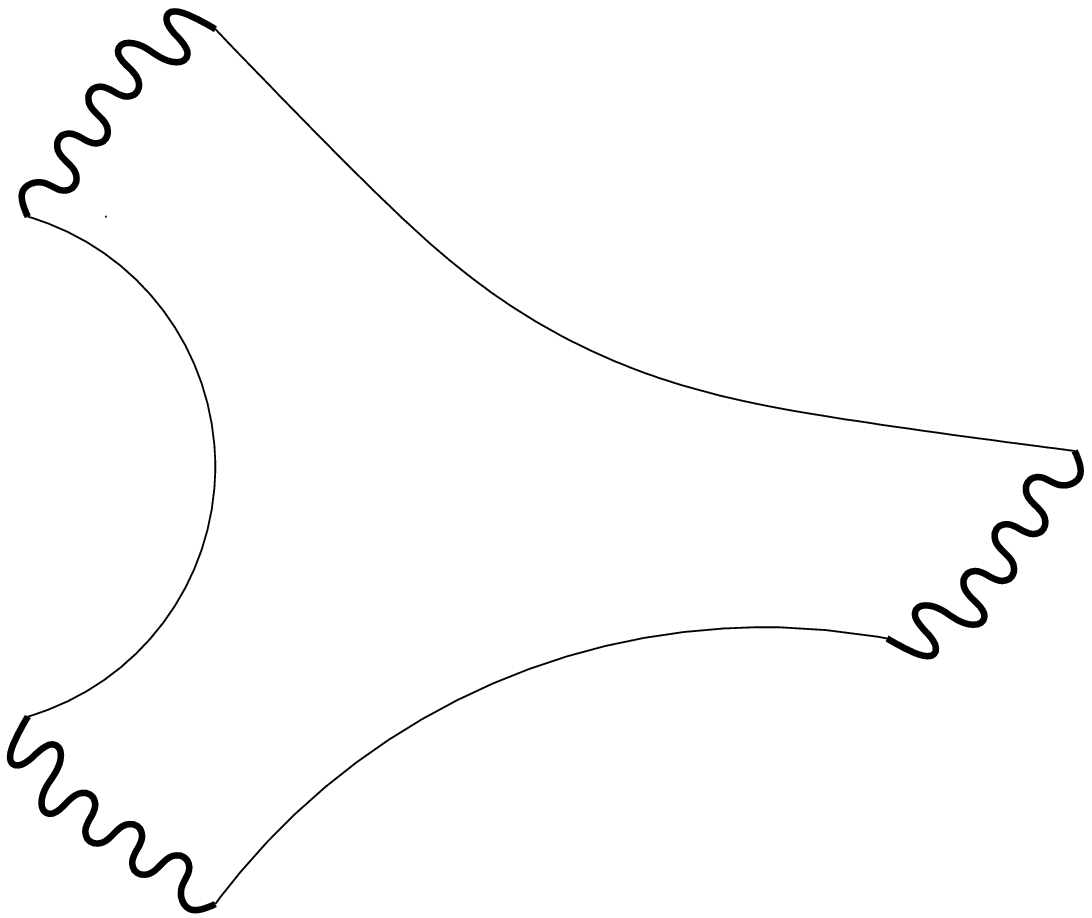,height=2.5cm} &
(c) \epsfig{file=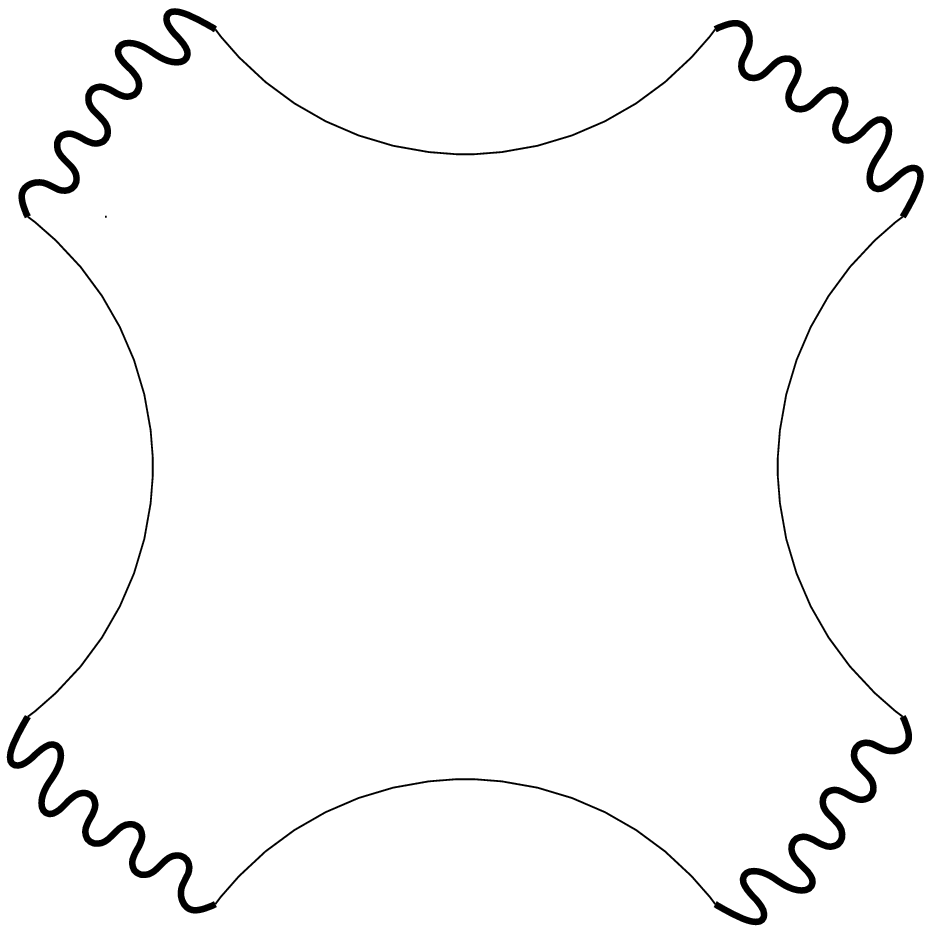,height=2.8cm} &
(e) \epsfig{file=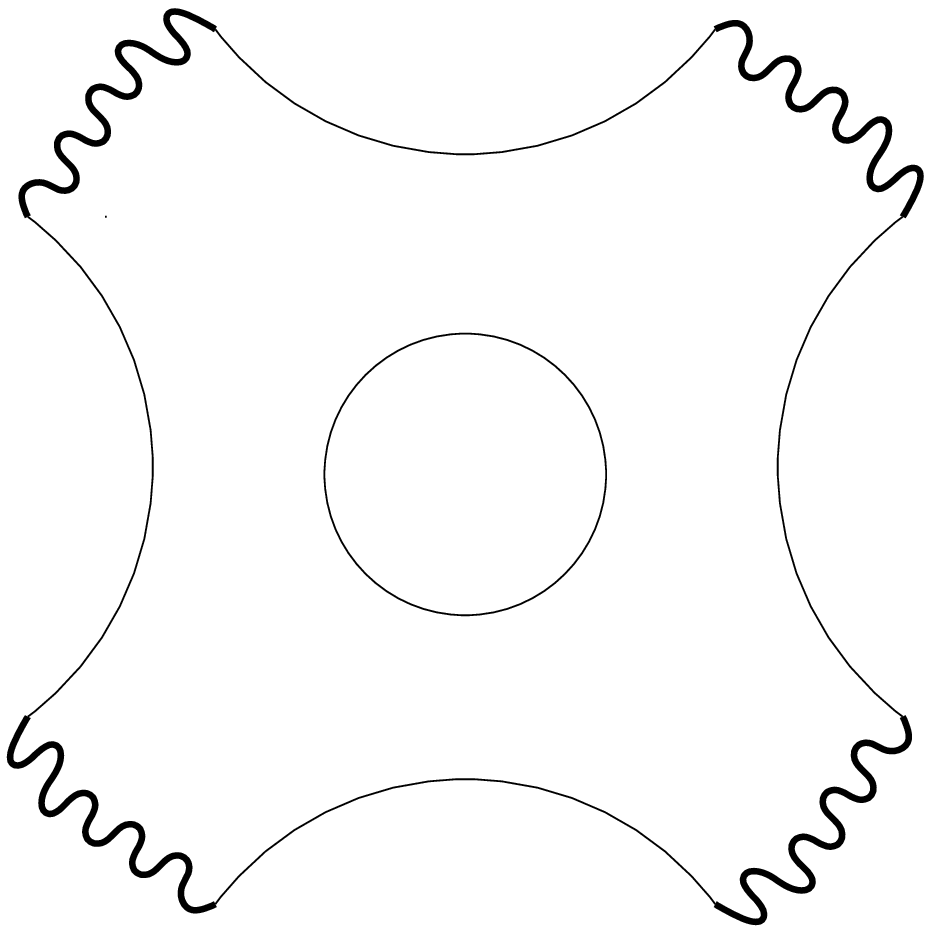,height=2.8cm} \\ 
(b) \epsfig{file=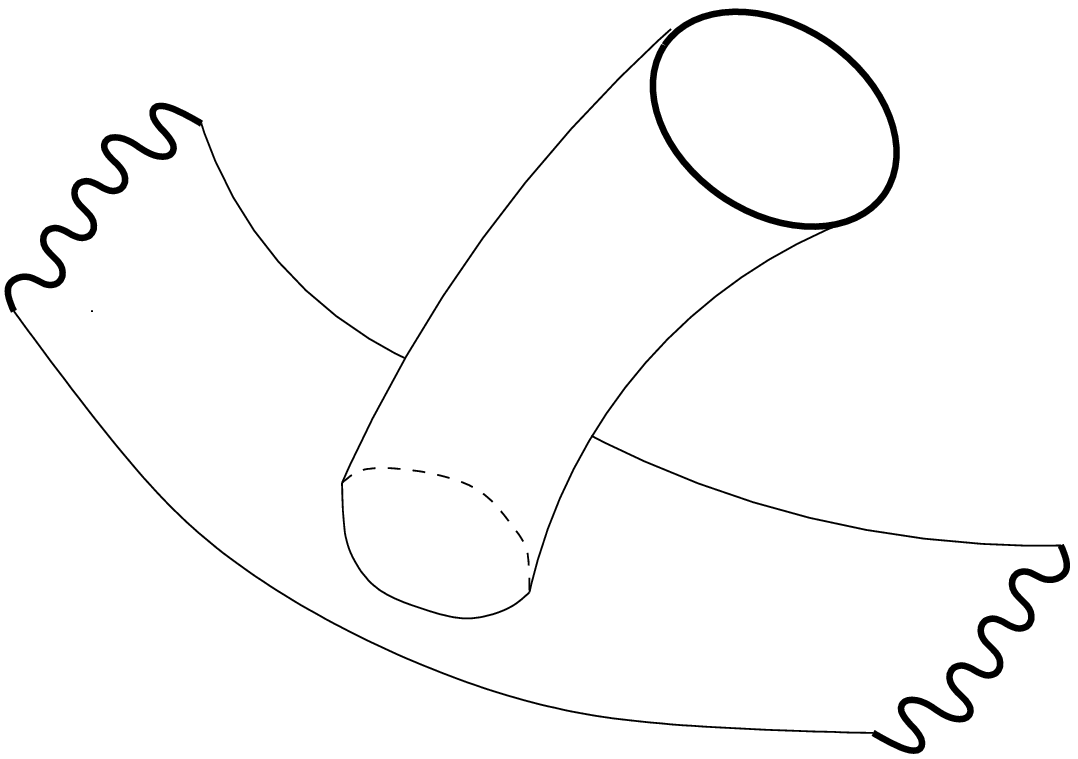,height=2cm} &
(d) \epsfig{file=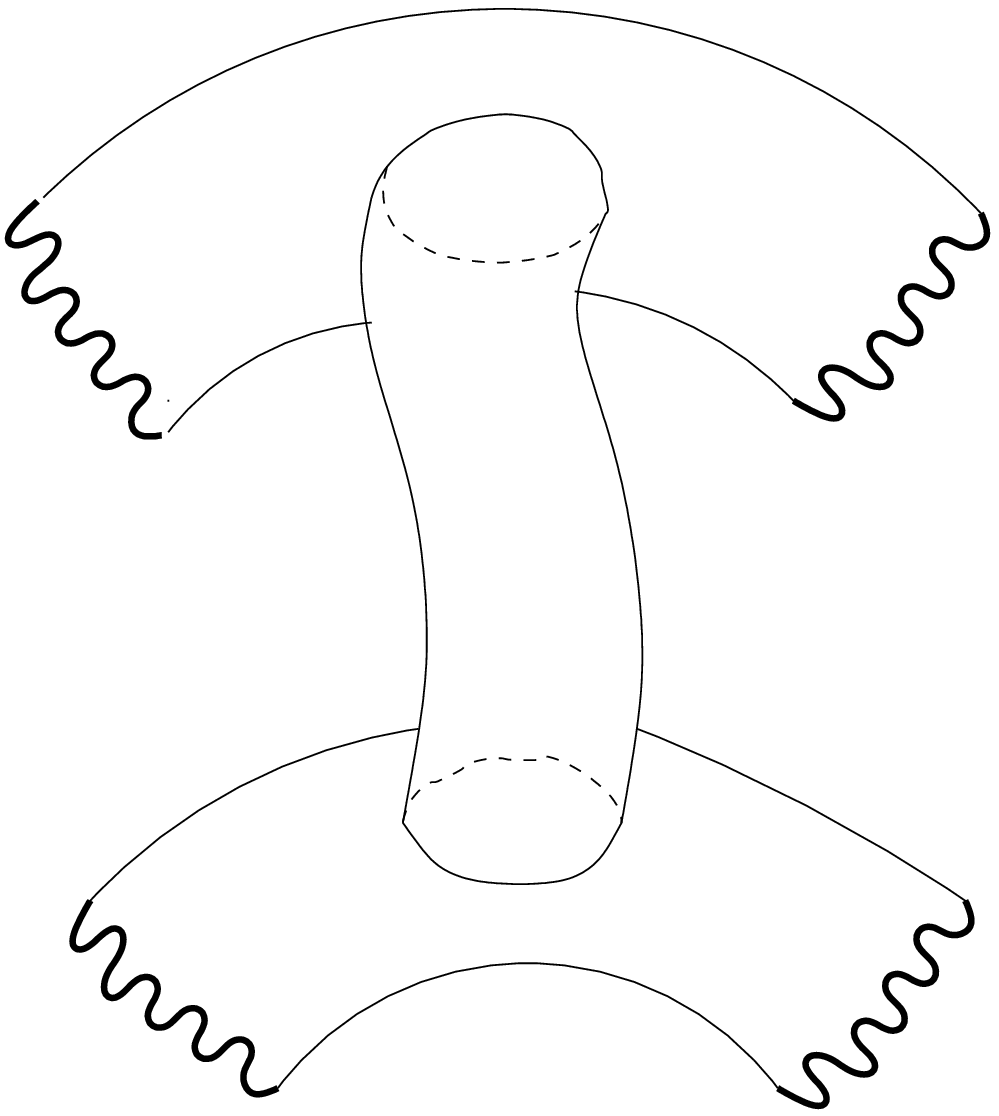,height=3cm} & 
\end{tabular}
\end{center}
\caption{
String graphs for 
(a) ${\cal O}(g)$ vertex of three open strings,
(b) ${\cal O}(\kappa)$ vertex of two open strings and one closed string,
(c) ${\cal O}(g^2)$ tree-level 2-body scattering of open strings (no gravity),
(d) ${\cal O}(\kappa^2)$ tree-level 2-body scattering with only gravity,
(e) ${\cal O}(g^4)$ loop-level 2-body scattering of open strings, 
    topological equivalent to (d) and hence $\kappa^2={\cal O}(g^4)$,
    meaning that (d) is $g^2$ smaller than (c).
\label{Fig1}}
\end{figure}

Now, as emphasized in \cite{Accomando:2000sj,Cullen:2000ef}, 
the exchange amplitude of a closed string has an order $g^2$ suppression 
versus the exchange of an open string, {\em in string perturbation theory}. 
This is illustrated in Fig.~1. In consequence, processes that receive
sizeable contributions from SR and Kaluza-Klein excitations of the 
graviton and also from SR excitations of the gauge bosons will 
be dominated by the second ones, for $s\lsim M^2_S/\alpha$.\footnote{
When the amplitude ${\cal A}_{\rm gauge}\sim\alpha s/M^2_S\lsim 
{\cal A}_{\rm grav}\sim \alpha^2s^2/M^6_S$, namely for $\alpha s \gsim M^2_S$,
gravity dominates and non-perturbative effects become important.}
That is, {\em gravity is subleading versus
gauge interactions} and hence the reactions are dominated at tree level
by open-string diagrams.
This is a generic feature in models of higher 
dimensional gravity embedded in a weakly-coupled string theory. 

\section{Phenomenology of TeV strings}

Cullen, Perelstein and Peskin introduced a TeV-scale string model for QED in 
\cite{Cullen:2000ef}. It contains electrons and photons at low energies 
and massive SR excitations above the string scale. 
Their results have been generalized in order to obtain string amplitudes 
for neutrino--quark elastic scattering \cite{Cornet:2001gy}, mediated 
in the SM by a $Z$ boson in the $t$--channel.

The model results from a simple embedding of the SM 
interactions into Type IIB string theory. It is assumed that 
the 10--d space of the theory has 6 dimensions compactified on a torus with 
common periodicity $2\pi R$,
and that $N$ coincident D3-branes (4--dimensional hypersurfaces
where open strings may end) are stretched out in the 4 extended
dimensions. Such configuration has N=4 supersymmetry and $U(N)$
gauge symmetry. One assumes that 
the extra symmetry of the massless string modes 
can be eliminated by an appropriate orbifold projection,
resulting an acceptable model with (at least) the SM
fields. The parameters of this theory would be the
string scale $M_S=\alpha'^{-1/2}\approx 1$ TeV and the 
dimensionless gauge coupling constant $g$, 
unified at $M_S$. Proposals for splitting these couplings
can be found in \cite{Ibanez:2000pw}. For more general 
D-brane models see \cite{Antoniadis:2000jv} and references therein.

A tree-level amplitude of open string states 
on a D-brane is given 
\cite{Cullen:2000ef,Hashimoto:1996kf} as a sum of ordered 
amplitudes multiplied by 
Chan-Paton traces. For the processes under study we have
\begin{eqnarray}
{\cal A}(1,2,3,4)&=&
g^2 \ \s(s,t) \ F^{1243}(s,t,u) \ 
\tr [t^1t^2t^4t^3+t^3t^4t^2t^1] \nonumber \\
&+& g^2 \ \s(s,u) \ F^{1234}(s,u,t) \ 
\tr [t^1t^2t^3t^4+t^4t^3t^2t^1] \nonumber \\
&+& g^2 \ \s(t,u) \ F^{1324}(t,u,s) \ 
\tr [t^1t^3t^2t^4+t^4t^2t^3t^1].
\label{string0}
\end{eqnarray}
In this expression, 
\begin{equation}
\s(s,t)={\Gamma(1-\alpha' s) \Gamma(1-\alpha' t)\over
\Gamma(1-\alpha' s-\alpha' t)}
\end{equation}
is basically the Veneziano amplitude \cite{Veneziano:1968yb}; 
$(1,2,3,4)$ label the external particles involved in the process:
$(\nu^{in}_L, u^{in}_L, \nu^{out}_L, u^{out}_L)$ for instance;
the Chan-Paton factors $t^a$ are representation
matrices of $U(N)$; and 
$F^{abcd}(s,t,u)$ is a factor depending
on the vertex operators for the external states and
their ordering.
In our case all the vertex operators correspond to (massless) Weyl
spinors of helicity (directed inward) $+$ or $-$, giving
\begin{eqnarray}
F^{-++-}(s,t,u)&=&-4{t\over s}; \nonumber \\
F^{-+-+}(s,t,u)&=&-4{u^2\over st}
=4 \left( {u\over s}+{u\over t}\right);\nonumber \\
F^{--++}(s,t,u)&=&-4{s\over t}.
\end{eqnarray}
Therefore the amplitude for the process $\nu_L u_L\to\nu_L u_L$ reads
\begin{equation}
{\cal A}=
-4 g^2 \left[ {s\over t} \; \s(s,t) \; T_{1243} + 
{s\over u} \; \s(s,u) \; T_{1234} + 
{s^2\over tu} \; \s(t,u) \; T_{1324} \right],
\label{ampl0}
\end{equation}
with $T_{abcd}$ the Chan-Paton traces.
To understand the phenomenological consequences of this amplitude
let us start with the limit $s,t\rightarrow 0$. Since $\Gamma(1)=1$,
we have all the Veneziano factors $\s (0,0)=1$. The amplitude
expresses then the exchange of massless vector modes in the $t$-- and 
the $u$--channels. The former would correspond to the $Z$ gauge 
boson, whereas the field exchanged in the $u$--channel is in the
$({\bf \overline 3}, {\bf 1})$ and/or the $({\bf \overline 3}, {\bf 3})$
representations of 
$SU(3)_C\times SU(2)_L$ and has electric charge $Q=-2/3$. 
We are interested, however, in models that {\em reproduce the SM result at
low energies}, with no massless leptoquarks. We obtain this limit 
if the Chan-Paton factors assigned to
$u_L$ and $\nu_L$ are such that
$T_{1243}-T_{1324}=-{1\over 10}$  and $T_{1234}=T_{1324}$,
where we have used $\sin^2\theta_W=3/8$, implied by gauge-coupling 
unification. 

In terms of $T_{1234}\equiv-a/10$ the amplitude (\ref{ampl0}) becomes
\begin{equation}
{\cal A}=
{2\over 5} g^2 \left[{s\over t} \left[ \left( 1+a \right)
\s(s,t) - a \s(t,u) \right] 
+ {s\over u} a\left[ \s(s,u) - \s(t,u) \right] \;\right].
\label{ampl1}
\end{equation}
At low $s$ this amplitude is 
${\cal A}_0\approx (2/5) g^2 {s/ t}$ and corresponds to the
exchange of a $Z$ boson in the $t$--channel.
The $Z$ is then a massless SR mode that 
acquires its mass $M_Z$ only through the Higgs 
mechanism. We shall neglect the corrections of order 
$M_Z^2/M_S^2$ that may affect the massive SR modes. 

As the energy increases the Veneziano factor $\s(s,t)$ gives a 
series of poles (at $1-\alpha' s=0,-1,-2,...$) and zeroes
(at $1-\alpha' s-\alpha' t=0,-1,-2,...$). It can be expressed as
\begin{equation}
\s(s,t)=\sum_{n=1}^{\infty} 
{\alpha' t+\alpha' s -1\over \alpha' t+ n - 1}\;
{\prod_{k=0}^{n-1}\; (\alpha' t + k)\over 
(\alpha' s -n)\; (n-1)!}\;.
\end{equation}
At $s=nM_S^2$ the amplitude describes the exchange
of a collection of resonances with the same mass 
and different spin (see below). Away from 
the poles the interference of resonances at different
mass levels produces the usual soft (Regge) behavior 
of the string in the ultraviolet. Obviously, these
resonances are not stable and at one loop will get an 
imaginary part in their propagator. When the total 
width of a resonance (which grows with its mass)
is similar to the mass difference with the 
resonance in the next level one cannot see resonances  
and interference effects dominate also at $s=nM_S^2$.

Let us first analyze the case with $a=0$ in Eq.~(\ref{ampl1}). 
The amplitude is just 
${\cal A}(\nu_L u_L \rightarrow \nu_L u_L)=
(2/ 5) g^2  (s/ t) \cdot \s(s,t)$.
Near the pole at $s=nM_S^2$,
\begin{equation}
{\cal A}_n\approx {2\over 5} g^2 \; {nM_S^4\over t}\; 
{(t/M_S^2)\; (t/M_S^2 +1)\cdot...\cdot 
( t/M_S^2 + n-1 ) \over 
(n-1)!\; (s - nM_S^2)}\;. 
\end{equation} 
This amplitude corresponds to the 
$s$--channel exchange of massive leptoquarks 
in the $({\bf 3}, {\bf 3})$ representation of 
$SU(3)_C\times SU(2)_L$ with 
electric charge $Q=2/3$. At each pole we have 
contributions of resonances with 
a common mass $\sqrt{n}M_S$ but different
spin, going from zero to the order of the 
residue $P_n(t)={\cal A}_n\cdot (s - nM_S^2)$. 
In this case the maximum
spin at the $n$ level is $J=n-1$.
To separate these contributions 
we first write the residue in terms of the
scattering angle $\theta$, with 
$t=-(nM_S^2/ 2)(1-\cos \theta)$.
Then we express $P_n(\theta)$ as a linear combination of
the $d$--functions (rotation matrix elements):
\begin{equation}
P_n(\theta)= {2\over 5} g^2 nM_S^2 \;
\sum_{J=0}^{n-1} \alpha^J_n\;
d^J_{0,0}(\theta)\;.
\end{equation}
The coefficient $\alpha^J_n$ gives the contribution
to our amplitude of a leptoquark $X^J_n$
of mass $nM_S^2$
and spin $J$. For example, at the first SR level
we find a scalar resonance with $\alpha^0_1=1$, 
at $s=2M_S^2$ there is a single vector resonance
with $\alpha^1_2=1$, whereas at 
$s=3M_S^2$ there are modes of spin 
$J=2$ ($\alpha^2_3=3/4$) and $J=0$ ($\alpha^0_3=1/4$).

The general case with $a\not= 0$ is completely analogous,
with resonant contributions from the terms 
proportional to $\s(s,t)$ and $\s(s,u)$.  
Taking $u=-(nM_S^2/ 2)(1+\cos \theta)$
and expressing again the residue in terms of
$d$--functions we find the same type of resonances
but with different $\alpha^J_n$ coefficients: 
$\alpha^0_1=1+2a$, $\alpha^1_2=1$, 
$\alpha^2_3=3(1+2a)/4$ and $\alpha^0_3=1(1+2a)/4$.  

\section{Application to the $\nu-$nucleon scattering}

From the resonant amplitude 
$\nu_L u_L\rightarrow X^J_n \rightarrow \nu_L u_L$ we can now obtain
the partial width 
$\Gamma^J_n\equiv \Gamma ( X^J_n \rightarrow \nu_L u_L)$:
\begin{equation}
\Gamma^J_n={g^2\over 40\pi}\; {\sqrt{n} M_S\; 
| \alpha^J_n | \over 2J+1}\;.
\end{equation}
Notice that for a given spin $J$, the variation with $n$ of 
$\alpha^J_n$ gives the {\it running} of the coupling with
the energy. We obtain numerically that the coupling of
heavier resonances decreases like the power law 
$\alpha^J_n\approx 1/n$.

The partial width $\Gamma^J_n$ can be used to obtain the 
cross section 
$\sigma^J_n(\nu_L u_L)\equiv \sigma (\nu_L u_L\rightarrow X^J_n)$ 
in the narrow-width approximation (NWA):
\begin{equation}
\sigma^J_n(\nu_L u_L)={4\pi^2\;\Gamma^J_n\over \sqrt{n} M_S}\;(2J+1)\;
\delta (s-nM_S^2)\;.
\end{equation}
At each mass level $n$ there is a tower of resonances of integer 
spin $J$ from 0 to $n-1$. We find a sum rule for the production 
rate $\sigma_n(\nu_L u_L)\equiv \sum_J \sigma^J_n(\nu_L u_L)$ 
of any of these resonances:
\begin{eqnarray}
\sigma_n (\nu_L u_L)&=&\left\{ \begin{array}{ll} 
\displaystyle {2\over 5}\;{\pi g^2\over 4}\;(1+2a)\;
\delta (s-nM_S^2)&{\rm for}\;\;n\;\;{\rm odd}
\vspace{0.2truecm} \\
\displaystyle {2\over 5}\;{\pi g^2\over 4}\;
\delta (s-nM_S^2)&{\rm for}\;\;n\;\;{\rm even}.
\end{array} \right.
\label{cs1}
\end{eqnarray}
Eq.~(\ref{cs1}) is equivalent (for $a=0$) to the production rate 
of a single resonance of mass $\sqrt{n} M_S$ and coupling 
$(2/5) g^2$ \cite{Doncheski:1997it}. This is a 
very interesting result: the coupling of heavier SR modes
decreases quadratically with the energy, but the number 
of modes (and the highest spin) 
at each mass level $n$ grows also quadratically making
$\sum_J \alpha^J_n$ a constant independent of $n$.

In the NWA 
$\sigma(\nu_L u_L) \equiv \sum_{n,J} \sigma(\nu_L u_L\rightarrow 
X^J_n \rightarrow {\rm anything})$ is then 
$\sigma(\nu_L u_L)=\sum_n \sigma_n(\nu_L u_L)$. In this limit the 
cross section is proportional to a collection of delta functions 
and thus all interference effects are ignored. This is a good
approximation as far as the total width of a resonance
is smaller than the mass difference with the next resonance
of same spin. Although the coupling (and any partial width)
decreases with the mass, the total width of
heavier resonances grows due to the larger number
of decay modes that are kinematically allowed. We estimate
that contributions to $\sigma(\nu_L u_L)$ from
modes beyond $n\cdot (g^2/4\pi) \approx 1$ 
are a continuum. In this regime any cross section goes
to zero exponentially at fixed angle ($s$ large, $t/s$ fixed)
and like a power law at small angles ($s$ large, $t$ fixed) 
\cite{pol98}. We neglect these contributions. We keep only 
resonant contributions from levels $n<n_{cut}= 50$. Our result 
depends very mildly on the actual value of $n_{cut}$.

To evaluate the total $\nu N$ cross section one also 
needs the elastic amplitudes $\nu_L d_L$, $\nu_L u_R$, $\nu_L d_R$
and $\nu_L \overline q_{L(R)}$. 
${\cal A}(\nu_L d_L \rightarrow \nu_L d_L)$ takes the same
form as the amplitude in Eq.~(\ref{ampl1}) with the changes 
$(2/5,a)\rightarrow (-3/5,a')$. 
The massive resonances exchanged in the 
$s$--channel are now an admixture of an 
$SU(2)_L$ singlet and a triplet. The singlet contribution
is required in $n$-even levels, otherwise an $SU(2)_L$ gauge 
transformation would relate the parameters $\alpha^J_n$ 
obtained here with the ones deduced from  
Eq.~(\ref{ampl1}).
In $n$-odd mass levels gauge invariance could be obtained
with no singlets for $a'=-(2+a)/3$.
The cross section $\sigma_n(\nu_L d_L)$ can be read from
Eq.~(\ref{cs1}) just by changing $(2/5,a)\rightarrow (3/5,a')$.

The calculation of amplitudes and cross sections for
$\nu_L\! \stackrel{_{(\_)}\ }{q_R}$ are completely analogous. 
We obtain
\begin{eqnarray}
\sigma_n(\nu_L u_R) &=&\left\{ \begin{array}{ll} 
\displaystyle {2\over 5}\;{\pi g^2\over 2}\;b\;
\delta (s-nM_S^2)&{\rm for}\;\;n\;\;{\rm odd} 
\vspace{0.2truecm} \\
0 & {\rm for}\;\;n\;\;{\rm even}.
\end{array} \right.
\label{cs3}
\end{eqnarray}
The cross sections $\sigma_n(\nu_L d_R)$, 
$\sigma_n(\nu_L \overline u_R)$ and
$\sigma_n(\nu_L \overline d_R)$ coincide with the
expression in Eq.~(\ref{cs3}) with the changes 
$(2/5,b)\rightarrow (1/5,b')$, 
$(2/5,b)\rightarrow (2/5,a)$ and 
$(2/5,b)\rightarrow (3/5,a')$, respectively.
For the left-handed antiquarks, 
$\sigma_n(\nu_L \overline u_L)$ and
$\sigma_n(\nu_L \overline d_L)$ can be read from
Eq.~(\ref{cs1}) by changing 
$(2/5,a)\rightarrow (2/5,b)$ and 
$(2/5,a)\rightarrow (1/5,b')$, respectively.

Now the total neutrino-nucleon cross section due
to the exchange of SR excitations can be very easily
evaluated. In terms of parton distribution functions
$q(x,Q)$ ($q=q_{L,R},\overline q_{L,R}$) in a 
nucleon $(N\equiv (n+p)/2)$ and the 
fraction of longitudinal momentum $x$, it is
\begin{equation}
\sigma (\nu_L N )= \sum_{n=1}^{n_{cut}} \sum_{q}
{\tilde \sigma_n (\nu_L q)\over nM_S^2}\; x\; q(x,Q),
\label{cs4}
\end{equation}
where $x=nM_S^2/s$, $s=2M_N E_\nu$, $Q^2=nM_S^2$ and 
$\tilde \sigma_n (\nu_L q)$ is the factor multiplying
the delta function in the cross section $\sigma_n (\nu_L q)$.

\begin{figure}[htb]
\epsfxsize=12cm
\epsfbox{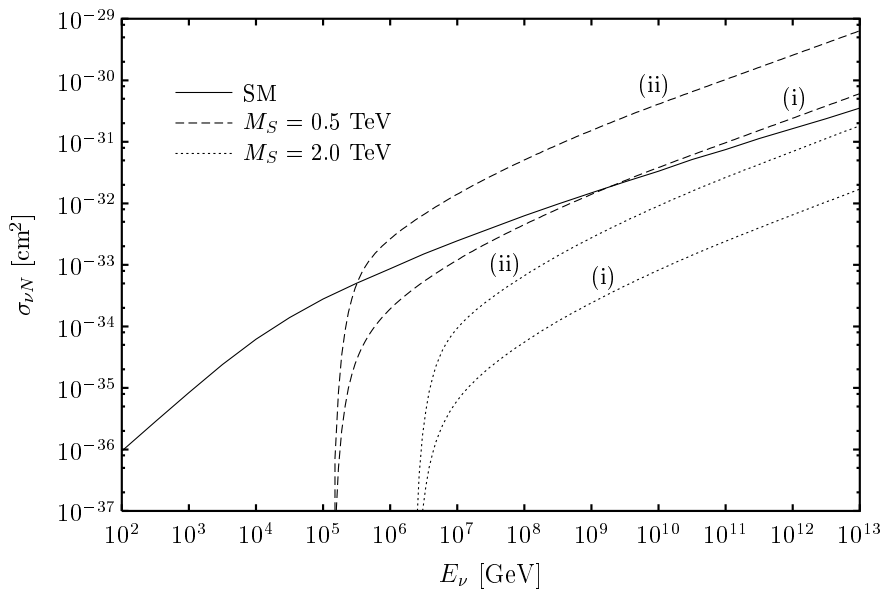}
\caption{
Neutrino-nucleon cross section versus the
incident neutrino energy $E_\nu$. The SM contribution
(solid) includes neutral and charged current interactions.
The SR contribution is for $M_S=0.5$ TeV (dashes)
and $M_S=2$ TeV (dots) for the cases (i)
$a=a'=b=b'=0$ and (ii) $a=a'=b=b'=5$.
\label{Fig2}}
\end{figure}

In Fig.~2 we plot the neutrino-nucleon cross section
at energies from $10^2$ to $10^{13}$ GeV for 
$M_S=0.5,2$ TeV. 
We have used
the CTEQ5 parton distributions in the DIS scheme 
\cite{Lai:2000wy} 
extended to $x < 10^{-5}$ with the methods in 
\cite{Gandhi:1996tf}. 
We show the SM cross section and 
the string corrections for 
$a=a'=b=b'$ equal 0 and 5 (notice that in the first case
there are no $s$--channel resonances mediating the 
$\nu_L q_R$ amplitude). The modes beyond 
$n_{cut}=50$ are not included, since there we expect that the narrow width 
approximation is poor. 

As a final comment, let us remark that the SM neutrino-nucleon 
interaction probes values of $x$ much below the current HERA DIS data 
($x^{\rm HERA}_{\rm min}\sim 10^{-4}$) for ultrahigh-energy neutrinos, 
$$
\langle x \rangle \approx \frac{1}{\sigma_{\rm SM}}\frac{G_F^2 M^2_W}{16\pi}
\frac{M_W}{\sqrt{s}}\approx 10 \left(\frac{E_\nu}{\rm GeV}\right)^{-0.8}
\approx 10^{-8} \mbox{ for } E_\nu=10^{11} \mbox{ GeV}.
$$
This is not the case for the process mediated by
SR resonances, where the probed values of $x$ are above
$$
x_{\rm min}\approx 0.5\times 10^6 \left(\frac{M_S}{\rm TeV}\right)^2
\left(\frac{E_\nu}{\rm GeV}\right)^{-1}
\approx 0.5\times 10^{-5} \mbox{ for } E_\nu=10^{11} \mbox{ GeV},
$$
taking $M_S=1$ TeV. 

\section{Conclusions}

Cosmic rays hit the nucleons in
the atmosphere with energies of up to $10^{11}$ GeV. If the
string scale is in the TeV range, these cosmic rays 
have the energy required to explore the fundamental 
theory and its interactions. In particular, ultrahigh-energy 
neutrinos are interesting since they can travel
long distances without losing a significant fraction of 
energy and are not deflected by magnetic fields. In addition, 
the SM interactions of a neutrino are much weaker than those 
of a quark or a charged lepton, which makes easier to see deviations 
due to new physics.

With this motivation we have analyzed the string
$\nu N$ cross section at energies much larger than
the fundamental scale $M_S\approx 1$ TeV. 
In a weakly-coupled string theory the process is given at tree level
by open-string graphs, whereas gravity effects appear as a one-loop 
correction.\footnote{Gravity effects in the context of weak-scale
string theories at high energies have been considered
in the Regge picture by \cite{Kachelriess:2000cb}.} We have fixed the 
arbitrary parameters of the model
imposing phenomenological constraints, namely,
the massless SR modes must account for the 
standard model particles. Four Chan-Paton traces remain as free parameters.
The massive SR modes include leptoquarks that 
resonantly mediate the process in the $s$--channel for
energies above the string scale.
The presence of massive leptoquarks 
is not a peculiarity of our toy model but a generic 
feature of any string model, and has to do with
$(s,t)$ and/or $(t,u)$ duality of the open-string amplitudes
(see {\it e.g.} Eq.~(\ref{ampl0})).

A very simple sum rule for the production rate of 
all the s--channel leptoquarks, with different spins 
in the same mass level $n$, makes possible
the calculation of the total $\nu N$ cross 
section in the narrow-width approximation.
The effect of these leptoquarks
is not just a correction of order 
$M^2_Z/M_S^2$ to the SM cross section, as one
would expect on dimensional grounds. 
Such SR excitations give a
contribution that can dominate for 
$M_S\approx 1$ TeV. 
However, for the expected flux of ultrahigh-energy neutrinos it 
seems unlikely that the cosmic ray events observed above
the GZK limit correspond to the 
decay of string resonances produced in $\nu N$ scattering. 
A similar conclusion has been recently drawn in \cite{Emparan:2001kf} 
for graviton-mediated $\nu N$ scattering and black hole 
production in TeV-gravity models.

Nevertheless, since neutrinos are very penetrating particles,
the enhancement of the cross-section may make possible the
detection of horizontal air showers produced by ultrahigh-energy CRs
in upcoming experiments \cite{Tyler:2001gt}. Furthermore,
in second-generation neutrino telescopes, able to detect
neutrinos in the TeV to PeV range, there is also a chance
to probe this and other models of TeV-scale quantum gravity
at more moderate energies \cite{Alvarez-Muniz:2001mk}.

\vspace{3mm}

It is my pleasure to thank the organizers of the XXV International
School of Theoretical Physics for their kind hospitality and
the stimulating and enjoyable atmosphere of the event. 
The author acknowledges F.~Cornet and M.~Masip for carefully reading 
the manuscript and a fruitful collaboration.

\end{document}